\documentclass[twocolumn]{aastex631}
\usepackage{graphicx,apjfonts}

\newcommand\uma{61\,UMa}
\newcommand\taucet{$\tau$\,Cet}
\newcommand\leo{88\,Leo}
\newcommand\rhocrb{$\rho$\,CrB}
\newcommand\sco{18\,Sco}
\newcommand\cygab{16\,Cyg\,A\,\&\,B}

\shorttitle{Constraints on Magnetic Braking}
\shortauthors{Metcalfe et al.}

\journalinfo{The Astrophysical Journal Letters}

\begin{document}

\title{\Large Constraints on Magnetic Braking from the G8 Dwarf Stars \uma\ and \taucet}

\author[0000-0003-4034-0416]{Travis S.~Metcalfe}
\affiliation{White Dwarf Research Corporation, 9020 Brumm Trail, Golden, CO 80403, USA}

\author[0000-0002-6192-6494]{Klaus G.~Strassmeier} 
\affiliation{Leibniz-Institut f\"ur Astrophysik Potsdam (AIP), An der Sternwarte 16, D-14482 Potsdam, Germany}

\author[0000-0002-0551-046X]{Ilya V.~Ilyin} 
\affiliation{Leibniz-Institut f\"ur Astrophysik Potsdam (AIP), An der Sternwarte 16, D-14482 Potsdam, Germany}

\author[0000-0002-4284-8638]{Jennifer L.~van~Saders} 
\affiliation{Institute for Astronomy, University of Hawai`i, 2680 Woodlawn Drive, Honolulu, HI 96822, USA}

\author[0000-0002-1242-5124]{Thomas R.~Ayres} 
\affiliation{Center for Astrophysics and Space Astronomy, 389 UCB, University of Colorado, Boulder, CO 80309, USA}

\author[0000-0002-3020-9409]{Adam J.~Finley} 
\affiliation{Department of Astrophysics-AIM, University of Paris-Saclay and University of Paris, CEA, CNRS, Gif-sur-Yvette Cedex F-91191, France}

\author[0000-0003-3061-4591]{Oleg Kochukhov} 
\affiliation{Department of Physics and Astronomy, Uppsala University, Box 516, SE-75120 Uppsala, Sweden}

\author[0000-0001-7624-9222]{Pascal Petit} 
\affiliation{Universit\'e de Toulouse, CNRS, CNES, 14 avenue Edouard Belin, 31400, Toulouse, France}

\author[0000-0001-5986-3423]{Victor~See} 
\affiliation{European Space Agency (ESA), European Space Research and Technology Centre (ESTEC), Keplerlaan 1, 2201 AZ Noordwijk, the Netherlands}

\author[0000-0002-3481-9052]{Keivan G.~Stassun} 
\affiliation{Vanderbilt University, Department of Physics \& Astronomy, 6301 Stevenson Center Lane, Nashville, TN 37235, USA}

\author[0000-0003-2490-4779]{Sandra V.~Jeffers} 
\affiliation{Max-Planck-Institut f\"ur Sonnensystemforschung, Justus-von-Liebig-weg 3, 37077, G\"ottingen, Germany}

\author[0000-0001-5522-8887]{Stephen C.~Marsden} 
\affiliation{Centre for Astrophysics, University of Southern Queensland, Toowoomba, Queensland, 4350, Australia}

\author[0000-0002-4996-6901]{Julien Morin} 
\affiliation{Centre national de la recherche scientifique (CNRS), Universit\'e de Montpellier, Place Eug\`ene Bataillon, 34095, Montpellier, France}

\author[0000-0001-5371-2675]{Aline A.~Vidotto} 
\affiliation{Leiden Observatory, Leiden University, PO Box 9513, 2300 RA, Leiden, The Netherlands}

\begin{abstract}

During the first half of their main-sequence lifetimes, stars rapidly lose angular 
momentum to their magnetized winds, a process known as magnetic braking. Recent 
observations suggest a substantial decrease in the magnetic braking efficiency when stars 
reach a critical value of the Rossby number, the stellar rotation period normalized by 
the convective overturn timescale. Cooler stars have deeper convection zones with longer 
overturn times, reaching this critical Rossby number at slower rotation rates. The nature 
and timing of the transition to weakened magnetic braking has previously been constrained 
by several solar analogs and two slightly hotter stars. In this Letter, we derive the 
first direct constraints from stars cooler than the Sun. We present new 
spectropolarimetry of the old G8 dwarf \taucet\ from the Large Binocular Telescope, and 
we reanalyze a published Zeeman Doppler image of the younger G8 star \uma, yielding the 
large-scale magnetic field strengths and morphologies. We estimate mass-loss rates using 
archival X-ray observations and inferences from Ly$\alpha$ measurements, and we adopt 
other stellar properties from asteroseismology and spectral energy distribution fitting. 
The resulting calculations of the wind braking torque demonstrate that the rate of 
angular momentum loss drops by a factor of 300 between the ages of these two stars 
(1.4--9~Gyr), well above theoretical expectations. We summarize the available data to 
help constrain the value of the critical Rossby number, and we identify a new signature 
of the long-period detection edge in recent measurements from the Kepler mission.

\end{abstract}


\section{Introduction}\label{sec1}

Seven years after the suggestion that weakened magnetic braking (WMB) might explain the 
anomalously rapid rotation of old Kepler field stars \citep{vanSaders2016}, the debate 
has shifted from whether this transition actually occurs, to the specific physical 
mechanisms that might drive it. Although the original sample included just 21 stars, the 
analysis was subsequently extended \citep{vanSaders2019} to reproduce the properties of 
34,000 Kepler field stars with measured rotation periods \citep{McQuillan2014}. The 
truncated distribution of rotation periods that led to the WMB hypothesis has now been 
confirmed using a sample of 94 stars with rotation rates measured from asteroseismology 
\citep{Hall2021}, and the predicted overdensity of stars with a range of ages near the 
long-period edge of the distribution has been confirmed \citep{David2022} using more than 
10,000 stars with precise effective temperatures from LAMOST spectra \citep{Xiang2019}.

Evidence quickly surfaced that the interruption to stellar rotational evolution was 
probably caused by an underlying transition in stellar magnetism. The influence of 
magnetic morphology on the rate of angular momentum loss from stellar winds 
\citep{Reville2015, Garraffo2015} led to the initial suggestion that WMB could be driven 
by a shift from a simple dipole to higher order fields \citep{vanSaders2016}. Preliminary 
support for this interpretation was identified \citep{Metcalfe2016} in measurements of 
the large-scale magnetic field strength and morphology from Zeeman Doppler imaging 
\citep{Petit2008}, and was reinforced by the associated changes in stellar activity 
cycles \citep{Metcalfe2017}. Targeted observations for an evolutionary sequence of two 
stars slightly hotter than the Sun \citep{Metcalfe2019, Metcalfe2021}, followed by a 
sequence of several solar analogs \citep{Metcalfe2022}, provided the first direct 
evidence of a magnetic morphology shift in older solar-type stars and placed new 
constraints on the relative importance of various contributions to the overall reduction 
in the wind braking torque. Such constraints may ultimately help to identify a 
corresponding transition in the stellar dynamo.

In this Letter, we present the first direct constraints on the nature and timing of WMB 
from stars that are cooler than the Sun. In Section~\ref{sec2}, we describe 
spectropolarimetric observations for an evolutionary sequence of two G8 dwarf stars, and 
we adopt other stellar properties largely from published sources. In Section~\ref{sec3}, 
we use the prescription of \cite{FinleyMatt2018} to estimate the wind braking torque for 
each star, confirming a dramatic decrease in the rate of angular momentum loss as seen 
previously for solar analogs and slightly hotter stars. Finally, we summarize the 
available constraints and we identify a new signature of the long-period detection edge 
in recent measurements from the Kepler mission.

\section{Observations}\label{sec2}

\subsection{Spectropolarimetry}\label{sec2.1} 

We observed \taucet\ on 2022 September 18 from the 2$\times$8.4\,m Large Binocular 
Telescope (LBT) using the Potsdam Echelle Polarimetric and Spectroscopic Instrument 
\citep[PEPSI;][]{Strassmeier2015}. The instrumental setup and data reduction methods were 
identical to those described in \cite{Metcalfe2019}, and we employed the least-squares 
deconvolution technique \citep[LSD;][]{Kochukhov2010} to derive precise mean intensity 
and polarization profiles. We obtained the line data required for the LSD analysis from 
the VALD database \citep{Ryabchikova2015}, adopting stellar atmospheric parameters from 
\cite{Brewer2016} and $v\sin i=0.4$~km~s$^{-1}$ from \cite{Saar1997}. Because \taucet\ 
has a 50~year record of constant chromospheric activity \citep{Baum2022} and appears to 
have a nearly pole-on orientation that minimizes rotational modulation \citep{Gray1994}, 
we assume that our snapshot observation is representative of the mean stellar activity 
level. The value of $\log R'_{\rm HK}$ for \taucet\ is above recent solar minimum levels 
\citep{Egeland2017}, although the actual value may be slightly lower due to its sub-solar 
metallicity \citep{SaarTesta2012}. There is no observational evidence to suggest that the 
low activity level of \taucet\ represents a magnetic grand minimum, it appears similar to 
other constant activity stars like \rhocrb\ and \cygab.

\begin{figure}[t]
 \centering\includegraphics[width=\columnwidth]{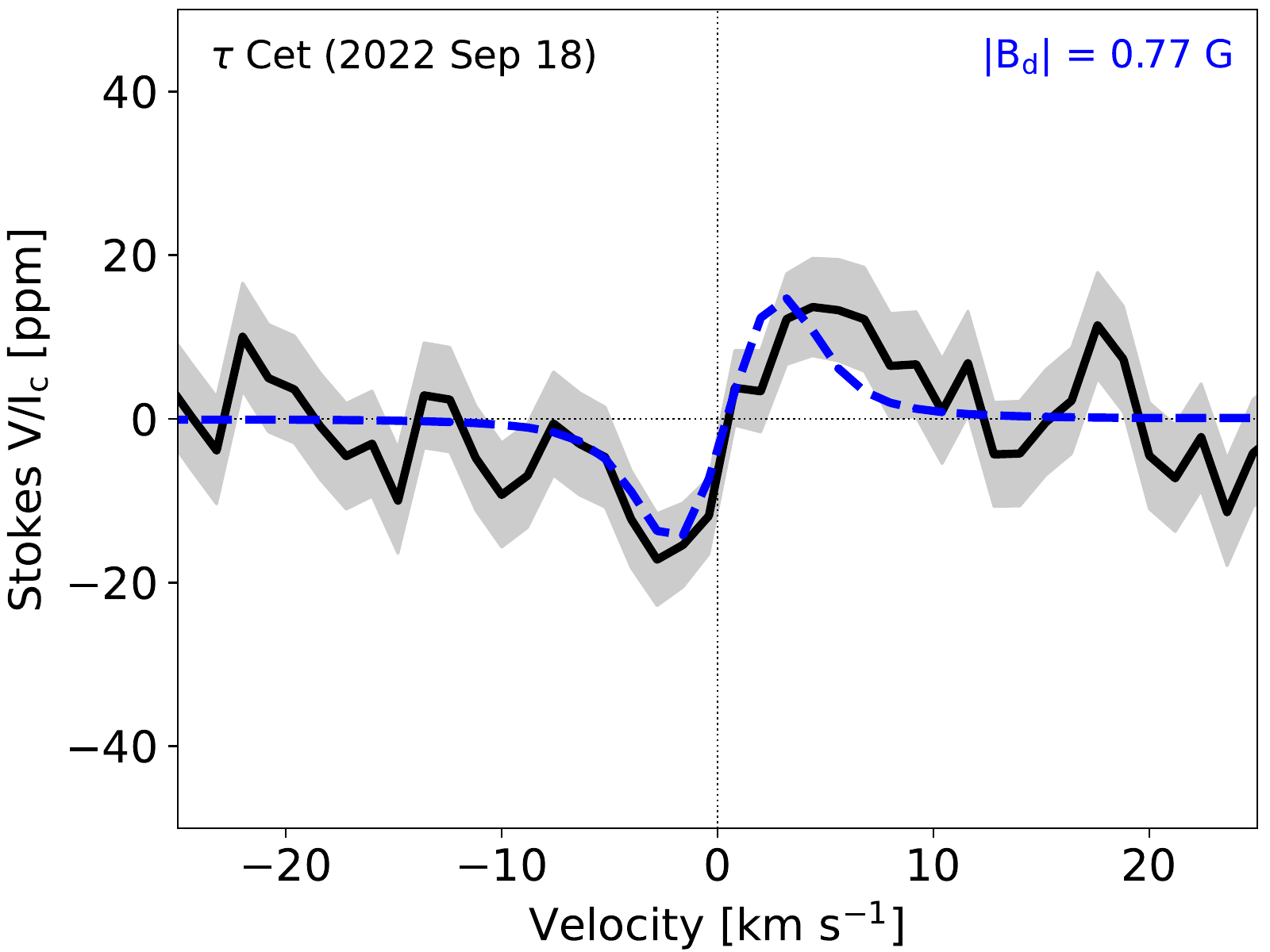}
 \caption{Stokes~$V$ polarization profile for \taucet\ from LBT observations on 2022 
September 18. The mean profile is shown as a black line with uncertainties indicated by 
the gray shaded area. The dashed blue line is an axisymmetric model profile assuming 
dipole morphology with the inclination fixed at $i\!=\!20^\circ$.\label{fig1}}
 \end{figure}

By combining the information from 1,463 metal lines deeper than 5\% of the continuum, we 
obtained a mean Stokes~$V$ profile with an uncertainty of 5.8~ppm (see 
Figure~\ref{fig1}). The data yielded a statistically marginal detection of the circular 
polarization signature for \taucet, with a mean longitudinal magnetic field $\langle 
B_{\rm z} \rangle$=$-0.37\pm0.08$~G. Following \cite{Metcalfe2019} we modeled the line 
profile assuming an axisymmetric dipole magnetic field with the inclination fixed at 
$i\!=\!20^\circ$ (see Section~\ref{sec2.2}), yielding a formal best fit $B_{\rm d}=-0.77 
\pm 0.31$~G. Note that the derived magnetic field strength does not depend strongly on 
the inclination (e.g., $B_{\rm d}=-0.73$~G for $i=5^\circ$).

 \begin{deluxetable*}{lccc}
 \setlength{\tabcolsep}{28pt}
 \tablecaption{Stellar Properties of 61~UMa and $\tau$~Cet\label{tab1}}
 \tablehead{\colhead{}             & \colhead{\uma}    & \colhead{\taucet} & \colhead{Sources}}
 \startdata
  $T_{\rm eff}$ (K)                & $5502 \pm 78$     & $5333 \pm 78$     & 1    \\
  $[$M/H$]$ (dex)                  & $-0.07 \pm 0.07$  & $-0.44 \pm 0.07$  & 1    \\
  $\log g$ (dex)                   & $4.52 \pm 0.08$   & $4.60 \pm 0.08$   & 1    \\
  $B-V$ (mag)                      & $0.72$            & $0.72$            & 2    \\
  $\log R'_{\rm HK}$ (dex)         & $-4.546$          & $-4.958$          & 2    \\
  $P_{\rm rot}$ (days)             & $17$              & $34$              & 2    \\
  $|B_{\rm d}|$ (G)                & $11.5$            & $0.77$            & 3    \\
  $|B_{\rm q}|$ (G)                & $12.0$            & $\cdots$          & 3    \\ 
  $|B_{\rm o}|$ (G)                & $6.12$            & $\cdots$          & 3    \\ 
  $L_X$ ($10^{27}$~erg~s$^{-1}$)   & $26.9 \pm 0.6$    & $0.49 \pm 0.10$   & 4    \\
  Mass-loss rate ($\dot{M}_\odot$) & $9.6 \pm 0.6$     & $<\!0.1$          & 4, 5 \\ 
  Mass ($M_\odot$)                 & $0.94 \pm 0.06$   & $0.783 \pm 0.012$ & 4, 6 \\
  Radius ($R_\odot$)               & $0.845 \pm 0.015$ & $0.816 \pm 0.012$ & 4, 7 \\
  Luminosity ($L_\odot$)           & $0.588 \pm 0.007$ & $0.473 \pm 0.011$ & 4    \\
  Age (Gyr)                        & $1.4 \pm 0.2$     & $9.0 \pm 1.0$     & 8, 9 \\
  \hline
  Torque ($10^{30}$~erg)           & $7.7$             & $<\!0.026$        & 10   \\
 \enddata
 \tablerefs{(1)~\cite{Brewer2016}; (2)~\cite{Baliunas1996}; (3)~Section\,\ref{sec2.1};
 (4)~Section\,\ref{sec2.2}; (5)~\cite{Wood2018}; (6)~\cite{Teixeira2009}; 
 (7)~\cite{vonBraun2017}; (8)~\cite{Barnes2007}; (9)~\cite{Tang2011}; 
 (10)~Section\,\ref{sec3}}
 \vspace*{-12pt}
 \end{deluxetable*}
 \vspace*{-24pt}

To complement the new LBT observations with another star along an evolutionary sequence, 
we reanalyzed a Zeeman Doppler imaging (ZDI) map for the younger G8 star \uma\ 
\citep{See2019}, which was obtained near the minimum of its 4~yr activity cycle. This ZDI 
map was based on 21 Stokes~$V$ measurements obtained in 2008 with the NARVAL 
spectropolarimeter on the 2.03\,m T{\'e}lescope Bernard Lyot. \cite{Folsom2018} provide 
details of the ZDI inversion procedure, while \cite{Jardine2013} discuss the connection 
between the resulting magnetic fields at the surface and in the corona. The wind braking 
prescription of \cite{FinleyMatt2018} requires the polar strengths of axisymmetric 
dipole, quadrupole, and octupole magnetic fields ($B_{\rm d}, B_{\rm q}, B_{\rm o}$) as 
input, but the observed ZDI map contains both axisymmetric and non-axisymmetric 
components. We followed the procedure described in \cite{Metcalfe2022} to calculate the 
equivalent polar field strengths for use with the wind braking prescription. This 
procedure captures the radial dependence of the magnetic flux for all components of the 
field, which is what matters for angular momentum loss. The results of this analysis for 
\uma\ are shown in Table~\ref{tab1}. There may be slight inconsistencies between the 
observed ZDI map and the equivalent polar field strengths derived in this way, but 
\cite{Jardine2013} demonstrated that the differences arise from non-radial components of 
the field, which induce near-surface magnetic stresses without altering the source 
surface of the stellar wind. Consequently, they are unimportant for our estimation of the 
wind braking torque.

\subsection{Stellar Properties}\label{sec2.2} 

In addition to the magnetic field strength and morphology, the wind braking prescription 
of \cite{FinleyMatt2018} also depends on the mass-loss rate ($\dot{M}$), the rotation 
period ($P_{\rm rot}$), and the stellar radius and mass ($R, M$). For \taucet\ we adopted 
the upper limit on the mass-loss rate from \cite{Wood2018}, which was obtained directly 
from Ly$\alpha$ measurements. The rotation period was determined from time series 
observations of chromospheric activity by \cite{Baliunas1996}, while the radius was 
inferred from interferometry \citep{vonBraun2017} and the mass was adopted from 
asteroseismology \citep{Teixeira2009}. The stellar age was derived from an analysis of 
these same asteroseismic observations by \cite{Tang2011}. Our estimate of the inclination 
$i\!=\!20^\circ$ in Section~\ref{sec2.1} was calculated from the values of $v\sin i$, 
$P_{\rm rot}$, and $R$.

For \uma, we determined the X-ray luminosity ($L_X$) from the ROSAT All-Sky Bright Source 
Catalog \citep{Boller2016}, following the approach described in \cite{Ayres2022}. When 
combined with the radius inferred from the spectral energy distribution (SED), following 
the procedures described in \cite{Stassun2017,Stassun2018}, this led to an X-ray flux per 
unit surface area that could be used to estimate the mass-loss rate from the empirical 
relation $\dot{M}\propto F_X^{0.77}$ \citep{Wood2021}. The rotation period of \uma\ was 
determined from chromospheric activity measurements \citep{Baliunas1996}, while the mass 
was estimated from the SED radius using the empirical mass-radius relation from eclipsing 
binaries \citep{Torres2010}. The age of \uma\ was determined from the gyrochronology 
relation of \cite{Barnes2007}, which remains valid for younger stars.

\section{Wind Braking Torque}\label{sec3}

We now have all of the observational inputs that are required to estimate the wind 
braking torque using the prescription of 
\cite{FinleyMatt2018}\footnote{\url{https://github.com/travismetcalfe/FinleyMatt2018}}. 
For \uma\ we have the equivalent polar field strengths ($B_{\rm d}, B_{\rm q}, B_{\rm 
o}$) from our reanalysis of the published ZDI map. For \taucet\ we have the snapshot 
observation from LBT, which we model with an axisymmetric dipole field inclined at 
$i\!=\!20^\circ$ (below we assess the influence of adopting a different inclination or 
morphology). We estimate the mass-loss rate ($\dot{M}$) from the X-ray flux for \uma, but 
we infer it directly from Ly$\alpha$ measurements for \taucet\ (below we assess the 
influence of the higher mass-loss rate estimated from the X-ray flux for \taucet). The 
remaining parameters ($P_{\rm rot}, R, M$) do not produce substantial uncertainties in 
the torque. Using the parameter values listed in Table~\ref{tab1} for our fiducial 
models, we find that the wind braking torque decreases by a factor of 300 between the 
ages of \uma\ and \taucet\ (1.4--9~Gyr).

By changing the wind braking prescription one parameter at a time between the values for 
\uma\ and \taucet, we can evaluate the relative importance of various contributions to 
the overall decrease in the torque. We find almost equal contributions from the 
evolutionary change in mass-loss rate ($-92$\%) and the difference in magnetic field 
strength and morphology ($-91$\%), with smaller contributions from the evolutionary 
change in rotation period ($-50$\%) and differences in the stellar radius ($-10$\%) and 
mass ($+4$\%). Rotational evolution models use the Rossby number (Ro) to scale the 
mass-loss rate as $\dot{M} \sim \textrm{Ro}^{-2}$ and the magnetic field strength as $B 
\sim P^{0.5}_{\rm phot} / \textrm{Ro}$, where $P_{\rm phot}$ is the photospheric pressure 
\citep{vanSaders2013}, but they do not directly account for changes in magnetic 
morphology. Considering either WMB or standard spin-down, these models predict a decrease 
in the wind braking torque of less than a factor of 15, with a smaller contribution from 
the evolutionary change in the mass-loss rate ($-70$\% to $-82$\%), and a difference from 
the change in magnetic field strength ($-40$\% to $-52$\%) that suggests a substantial 
contribution from a shift in magnetic morphology.

The large decrease in wind braking torque between the ages of \uma\ and \taucet\ is 
robust against changes to our assumptions about the magnetic field and mass-loss rate. 
For example, \taucet\ has a debris disk with a measured inclination \citep[$i=35^\circ 
\pm 10^\circ$;][]{Lawler2014}. If we assume that the rotation axis of \taucet\ shares 
this same inclination, the inferred dipole field would be slightly stronger ($B_{\rm 
d}=-0.89$~G) and produce a corresponding increase in the wind braking torque 
(0.029$\times10^{30}$\,erg, $+14$\%). If we keep $i\!=\!20^\circ$ but model the 
Stokes~$V$ profile as an axisymmetric quadrupole or octupole field, enhanced geometric 
cancelation would lead us to infer a stronger field ($B_{\rm q}=-2.94$~G or $B_{\rm 
o}=-118.9$~G). For a quadrupole field this ultimately leads to a lower wind braking 
torque (0.023$\times10^{30}$\,erg, $-11$\%), because the stronger field does not 
compensate for the shorter effective lever-arm. By contrast, the much stronger octupole 
field would ultimately produce a higher torque (0.043$\times10^{30}$\,erg, $+67$\%) 
despite the much shorter effective lever-arm, although this scenario is incompatible with 
the measured chromospheric activity level. Finally, if we use the X-ray flux of \taucet\ 
to estimate its mass-loss rate (0.46~$\dot{M}_\odot$) following the same approach as for 
\uma, the wind braking torque would more than double (0.058$\times10^{30}$\,erg, 
$+129$\%). However, even if we adopt both the octupole field and the higher mass-loss 
rate for \taucet\ (contradicting the actual measurements of $\log R'_{\rm HK}$ and 
$\dot{M}$ shown in Table~\ref{tab1}), the wind braking torque would still be 50 times 
weaker than for \uma.

 \begin{figure}[t]
 \centering\includegraphics[width=\columnwidth]{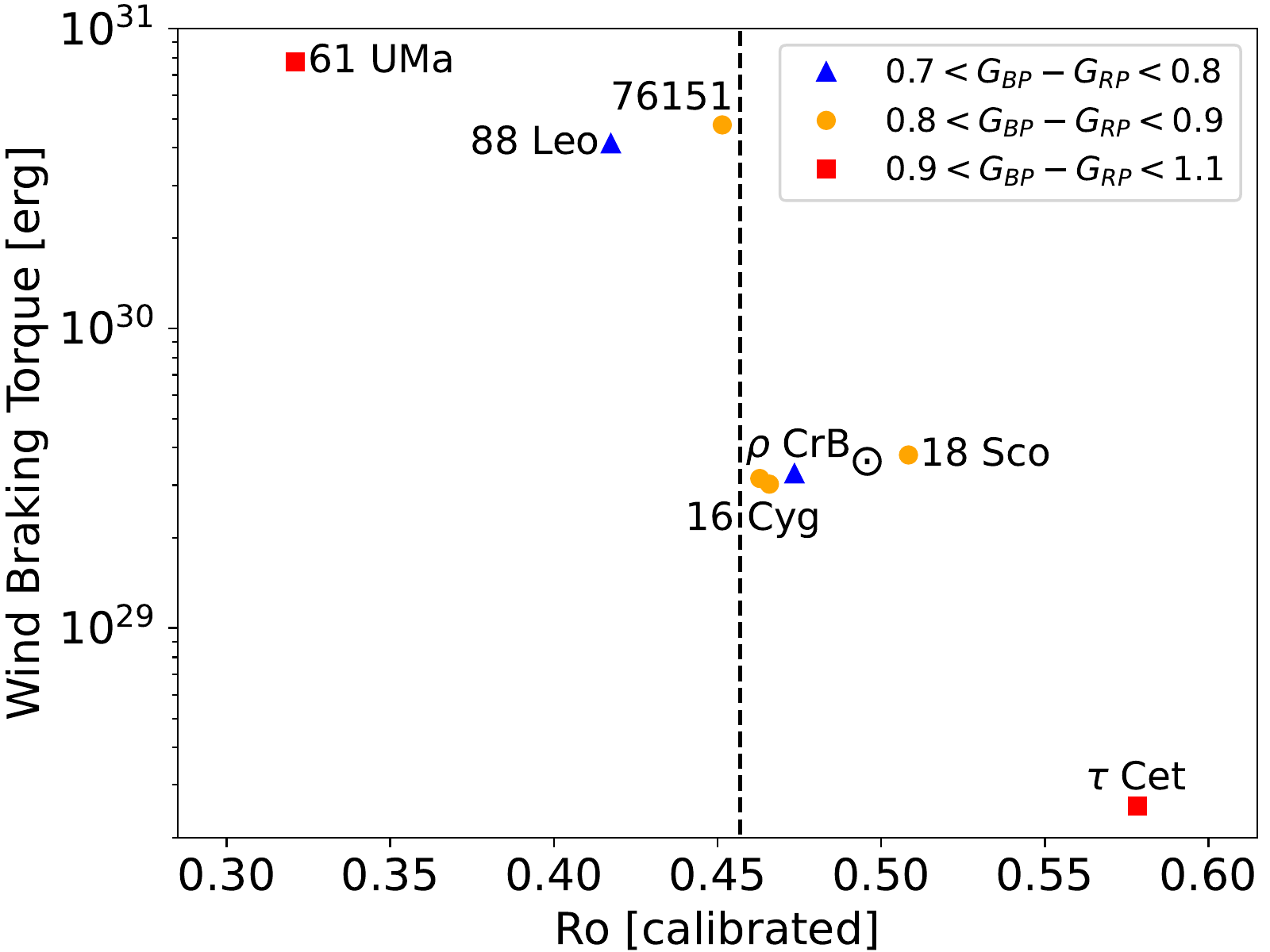}
 \caption{Evolution of the wind braking torque with Ro from the calibration of 
\cite{Corsaro2021}. Points are grouped by Gaia color, corresponding to solar analogs 
(yellow circles), and hotter (blue triangles) or cooler stars (red squares). The 
empirical constraint for Ro$_{\rm crit}$ on this scale is shown with a vertical dashed 
line.\label{fig2}}
 \end{figure}

We can assess the empirical value of the critical Rossby number (Ro$_{\rm crit}$) by 
plotting the available estimates of wind braking torque on a common scale. We use the 
asteroseismic calibration of \cite{Corsaro2021} for the convective overturn timescale to 
calculate Ro for each star in our sample based on the Gaia $G_{BP}-G_{RP}$ color. The 
results are shown in Figure~\ref{fig2} with points grouped by Gaia color, corresponding 
to solar analogs (yellow circles), and hotter (blue triangles) or cooler stars (red 
squares). Our empirical constraint on the critical value of Ro on this scale (Ro$_{\rm 
crit}=0.457 \pm 0.006$) is shown with a vertical dashed line. Our new estimates of the 
wind braking torque for \uma\ and \taucet\ extend the range of Ro sampled by our 
spectropolarimetric targets, and show the highest and lowest torques of the ensemble. By 
construction, the more active stars in our sample (\uma, \leo, HD\,76151) are all below 
Ro$_{\rm crit}$ while the less active stars (\cygab, \rhocrb, \sco, \taucet) are all 
above it. The Rossby numbers of HD\,76151 (Ro\,=\,0.451) and 16\,Cyg\,A (Ro\,=\,0.463) 
currently define the empirical constraint on the value of Ro$_{\rm crit}$ shown above, 
falling slightly below the solar value (Ro$_\odot$\,=\,0.496) where we plot the solar 
wind braking torque determined by \cite{Finley2018}.

\section{Discussion}\label{sec4}

Using spectropolarimetric constraints on the large-scale magnetic field, we have 
demonstrated that the rate of angular momentum loss due to stellar winds decreases by a 
factor of 300 between the ages of two stars that are cooler than the Sun, dominated by 
contributions from both the mass-loss rate and the magnetic field strength and 
morphology. Relative to previous comparisons for solar analogs and slightly hotter stars, 
there is a larger evolutionary gap between \uma\ and \taucet, which would naturally lead 
to a more substantial decrease in the strength of magnetic braking across Ro$_{\rm 
crit}$. On the other hand, the estimated wind braking torque for \taucet\ could be even 
smaller because the mass-loss rate from \cite{Wood2018} is an upper limit. For the solar 
analogs \cygab, \cite{Metcalfe2022} estimated a factor of 15--16 decrease in the wind 
braking torque relative to HD\,76151, while for the slightly hotter star \rhocrb, 
\cite{Metcalfe2021} estimated a factor of 13 decrease relative to \leo. However, these 
estimates all adopted upper limits on the magnetic field strengths from statistical 
non-detections of the Stokes~$V$ signatures, so the actual decreases could be larger. Our 
marginal detection of the Stokes~$V$ signature for \taucet\ was facilitated by lower 
noise levels and by the nearly pole-on orientation, which minimizes geometric cancelation 
for an axisymmetric dipole magnetic field.

\begin{figure}[t]
 \centering\includegraphics[width=\columnwidth]{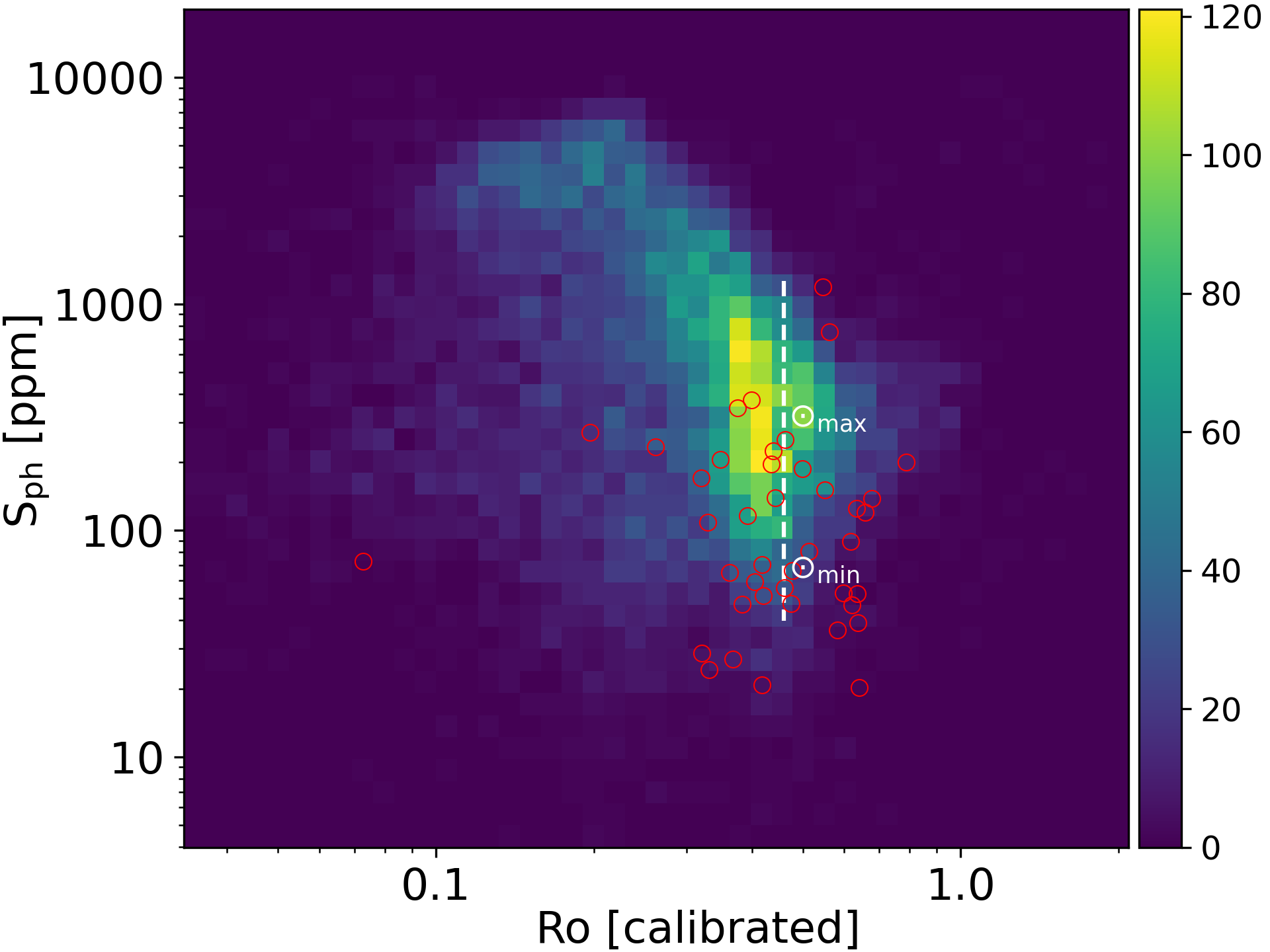}
 \caption{Rotation-activity relation for stars published in \cite{Corsaro2021}, with the 
critical value of Ro from Section~\ref{sec3} shown as a vertical dashed line. Bright 
asteroseismic targets are shown with red circles. The highest density of stars (yellow 
and green) exhibit a broad range of activity levels at nearly constant Ro, which may 
correspond to the long-period edge of the Kepler sample.\label{fig3}}
 \end{figure}

From a sample of 40,000 Kepler targets with measured rotation periods, \cite{Corsaro2021} 
plotted a photometric proxy for stellar activity \citep[$S_{\rm ph}$;][]{Mathur2014} 
against the newly calibrated values of Ro (see Figure~\ref{fig3}). Despite the vertical 
dispersion at all values of Ro due to activity cycles and inclination effects, the 
highest density of stars in their sample (yellow and green) appears at Ro slightly below 
the solar value, showing a broad range of stellar activity levels \citep{Santos2019, 
Santos2021} at nearly constant Ro. This feature may correspond to the long-period edge of 
the Kepler sample \citep{McQuillan2014, vanSaders2019}, where stellar activity gradually 
declines with age at roughly constant rotation period during the second half of 
main-sequence lifetimes \citep{Metcalfe2016, Metcalfe2017}. However, the relatively 
uniform distribution of bright asteroseismic targets (red circles) across our empirical 
value of Ro$_{\rm crit}$ and toward low values of $S_{\rm ph}$ suggests that the 
detection of rotational modulation may simply be less efficient at higher Ro and lower 
$S_{\rm ph}$ \citep{vanSaders2019, Masuda2022}.

Future spectropolarimetry promises to extend our measurements to even cooler K-type 
stars, and provide additional constraints on the empirical value of Ro$_{\rm crit}$. 
Snapshot observations at LBT are already planned for HD\,103095 and HD\,166620. Although 
HD\,103095 is an early K-type star, its extremely low metallicity 
\citep[$\mathrm{[M/H]}$=$-1.16$;][]{ValentiFischer2005} gives it the shallower convection 
zone of a late G-type star similar to \taucet. With an activity cycle of 7.3 years and a 
measured rotation period of 31 days \citep{Baliunas1995, Baliunas1996}, it will sample 
intermediate conditions (Ro\,=\,0.542) between \uma\ and \taucet\ and probe how the 
driving mechanisms of WMB change over time. Observations of HD\,166620 will complement 
our existing measurements of the active star 40\,Eri, probing the magnetic morphology of 
a K-type star that recently entered a grand magnetic minimum \citep{Baum2022, Luhn2022}. 
Archival ZDI maps of younger K-type stars like $\epsilon$\,Eri \citep{Jeffers2014} and 
61\,Cyg\,A \citep{BoroSaikia2016} from the BCool collaboration will provide additional 
context for our understanding of how magnetic braking shapes the lives of other Sun-like 
stars.

\vspace*{12pt}
Special thanks to Steve Saar, \^{A}ngela Santos, and Brian Wood for helpful exchanges. 
T.S.M.\ acknowledges support from the U.S.\ National Science Foundation (AST-2205919).
A.J.F.\ and A.A.V.\ have received funding from the European Research Council (ERC) under the European Union's Horizon 2020 research and innovation programme (grant agreements No 810218 WHOLESUN, and No 817540 ASTROFLOW).
O.K.\ acknowledges support by the Swedish Research Council (project 2019-03548) and the Royal Swedish Academy of Sciences.
V.S.\ acknowledges support from the European Space Agency (ESA) as an ESA Research Fellow.
The LBT is an international collaboration among institutions in the United States, Italy and Germany. LBT Corporation partners are: The University of Arizona on behalf of the Arizona Board of Regents; Istituto Nazionale di Astrofisica, Italy; LBT Beteiligungsgesellschaft, Germany, representing the Max-Planck Society, The Leibniz Institute for Astrophysics Potsdam, and Heidelberg University; The Ohio State University, and The Research Corporation, on behalf of The University of Notre Dame, University of Minnesota and University of Virginia.



\begin{thebibliography}{52}
\expandafter\ifx\csname natexlab\endcsname\relax\def\natexlab#1{#1}\fi

\bibitem[{{Ayres} \& {Buzasi}(2022)}]{Ayres2022}
{Ayres}, T., \& {Buzasi}, D. 2022, \apjs, 263, 41

\bibitem[{{Baliunas} {et~al.}(1996){Baliunas}, {Sokoloff}, \&
  {Soon}}]{Baliunas1996}
{Baliunas}, S., {Sokoloff}, D., \& {Soon}, W. 1996, \apjl, 457, L99

\bibitem[{{Baliunas} {et~al.}(1995){Baliunas}, {Donahue}, {Soon}, {Horne},
  {Frazer}, {Woodard-Eklund}, {Bradford}, {Rao}, {Wilson}, {Zhang}, {Bennett},
  {Briggs}, {Carroll}, {Duncan}, {Figueroa}, {Lanning}, {Misch}, {Mueller},
  {Noyes}, {Poppe}, {Porter}, {Robinson}, {Russell}, {Shelton}, {Soyumer},
  {Vaughan}, \& {Whitney}}]{Baliunas1995}
{Baliunas}, S.~L., {et~al.} 1995, \apj, 438, 269

\bibitem[{{Barnes}(2007)}]{Barnes2007}
{Barnes}, S.~A. 2007, \apj, 669, 1167

\bibitem[{{Baum} {et~al.}(2022){Baum}, {Wright}, {Luhn}, \&
  {Isaacson}}]{Baum2022}
{Baum}, A.~C., {Wright}, J.~T., {Luhn}, J.~K., \& {Isaacson}, H. 2022, \aj,
  163, 183

\bibitem[{{Boller} {et~al.}(2016){Boller}, {Freyberg}, {Tr{\"u}mper}, {Haberl},
  {Voges}, \& {Nandra}}]{Boller2016}
{Boller}, T., {Freyberg}, M.~J., {Tr{\"u}mper}, J., {Haberl}, F., {Voges}, W.,
  \& {Nandra}, K. 2016, \aap, 588, A103

\bibitem[{{Boro Saikia} {et~al.}(2016){Boro Saikia}, {Jeffers}, {Morin},
  {Petit}, {Folsom}, {Marsden}, {Donati}, {Cameron}, {Hall}, {Perdelwitz},
  {Reiners}, \& {Vidotto}}]{BoroSaikia2016}
{Boro Saikia}, S., {et~al.} 2016, \aap, 594, A29

\bibitem[{{Brewer} {et~al.}(2016){Brewer}, {Fischer}, {Valenti}, \&
  {Piskunov}}]{Brewer2016}
{Brewer}, J.~M., {Fischer}, D.~A., {Valenti}, J.~A., \& {Piskunov}, N. 2016,
  \apjs, 225, 32

\bibitem[{{Corsaro} {et~al.}(2021){Corsaro}, {Bonanno}, {Mathur},
  {Garc{\'\i}a}, {Santos}, {Breton}, \& {Khalatyan}}]{Corsaro2021}
{Corsaro}, E., {Bonanno}, A., {Mathur}, S., {Garc{\'\i}a}, R.~A., {Santos},
  A.~R.~G., {Breton}, S.~N., \& {Khalatyan}, A. 2021, \aap, 652, L2

\bibitem[{{David} {et~al.}(2022){David}, {Angus}, {Curtis}, {van Saders},
  {Colman}, {Contardo}, {Lu}, \& {Zinn}}]{David2022}
{David}, T.~J., {Angus}, R., {Curtis}, J.~L., {van Saders}, J.~L., {Colman},
  I.~L., {Contardo}, G., {Lu}, Y., \& {Zinn}, J.~C. 2022, \apj, 933, 114

\bibitem[{{Egeland} {et~al.}(2017){Egeland}, {Soon}, {Baliunas}, {Hall},
  {Pevtsov}, \& {Bertello}}]{Egeland2017}
{Egeland}, R., {Soon}, W., {Baliunas}, S., {Hall}, J.~C., {Pevtsov}, A.~A., \&
  {Bertello}, L. 2017, \apj, 835

\bibitem[{{Finley} \& {Matt}(2018)}]{FinleyMatt2018}
{Finley}, A.~J., \& {Matt}, S.~P. 2018, \apj, 854, 78

\bibitem[{{Finley} {et~al.}(2018){Finley}, {Matt}, \& {See}}]{Finley2018}
{Finley}, A.~J., {Matt}, S.~P., \& {See}, V. 2018, \apj, 864, 125

\bibitem[{{Folsom} {et~al.}(2018){Folsom}, {Bouvier}, {Petit}, {L{\`e}bre},
  {Amard}, {Palacios}, {Morin}, {Donati}, \& {Vidotto}}]{Folsom2018}
{Folsom}, C.~P., {et~al.} 2018, \mnras, 474, 4956

\bibitem[{{Garraffo} {et~al.}(2015){Garraffo}, {Drake}, \&
  {Cohen}}]{Garraffo2015}
{Garraffo}, C., {Drake}, J.~J., \& {Cohen}, O. 2015, \apj, 813, 40

\bibitem[{{Gray} \& {Baliunas}(1994)}]{Gray1994}
{Gray}, D.~F., \& {Baliunas}, S.~L. 1994, \apj, 427, 1042

\bibitem[{{Hall} {et~al.}(2021){Hall}, {Davies}, {van Saders}, {Nielsen},
  {Lund}, {Chaplin}, {Garc{\'\i}a}, {Amard}, {Breimann}, {Khan}, {See}, \&
  {Tayar}}]{Hall2021}
{Hall}, O.~J., {et~al.} 2021, Nature Astronomy, 5, 707

\bibitem[{{Jardine} {et~al.}(2013){Jardine}, {Vidotto}, {van Ballegooijen},
  {Donati}, {Morin}, {Fares}, \& {Gombosi}}]{Jardine2013}
{Jardine}, M., {Vidotto}, A.~A., {van Ballegooijen}, A., {Donati}, J.~F.,
  {Morin}, J., {Fares}, R., \& {Gombosi}, T.~I. 2013, \mnras, 431, 528

\bibitem[{{Jeffers} {et~al.}(2014){Jeffers}, {Petit}, {Marsden}, {Morin},
  {Donati}, \& {Folsom}}]{Jeffers2014}
{Jeffers}, S.~V., {Petit}, P., {Marsden}, S.~C., {Morin}, J., {Donati}, J.~F.,
  \& {Folsom}, C.~P. 2014, \aap, 569, A79

\bibitem[{{Kochukhov} {et~al.}(2010){Kochukhov}, {Makaganiuk}, \&
  {Piskunov}}]{Kochukhov2010}
{Kochukhov}, O., {Makaganiuk}, V., \& {Piskunov}, N. 2010, \aap, 524, A5

\bibitem[{{Lawler} {et~al.}(2014){Lawler}, {Di Francesco}, {Kennedy},
  {Sibthorpe}, {Booth}, {Vandenbussche}, {Matthews}, {Holland}, {Greaves},
  {Wilner}, {Tuomi}, {Blommaert}, {de Vries}, {Dominik}, {Fridlund}, {Gear},
  {Heras}, {Ivison}, \& {Olofsson}}]{Lawler2014}
{Lawler}, S.~M., {et~al.} 2014, \mnras, 444, 2665

\bibitem[{{Luhn} {et~al.}(2022){Luhn}, {Wright}, {Henry}, {Saar}, \&
  {Baum}}]{Luhn2022}
{Luhn}, J.~K., {Wright}, J.~T., {Henry}, G.~W., {Saar}, S.~H., \& {Baum}, A.~C.
  2022, \apjl, 936, L23

\bibitem[{{Masuda}(2022)}]{Masuda2022}
{Masuda}, K. 2022, \apj, 937, 94

\bibitem[{{Mathur} {et~al.}(2014){Mathur}, {Garc{\'\i}a}, {Ballot}, {Ceillier},
  {Salabert}, {Metcalfe}, {R{\'e}gulo}, {Jim{\'e}nez}, \&
  {Bloemen}}]{Mathur2014}
{Mathur}, S., {et~al.} 2014, \aap, 562, A124

\bibitem[{{McQuillan} {et~al.}(2014){McQuillan}, {Mazeh}, \&
  {Aigrain}}]{McQuillan2014}
{McQuillan}, A., {Mazeh}, T., \& {Aigrain}, S. 2014, \apjs, 211, 24

\bibitem[{{Metcalfe} {et~al.}(2016){Metcalfe}, {Egeland}, \& {van
  Saders}}]{Metcalfe2016}
{Metcalfe}, T.~S., {Egeland}, R., \& {van Saders}, J. 2016, \apjl, 826, L2

\bibitem[{{Metcalfe} {et~al.}(2019){Metcalfe}, {Kochukhov}, {Ilyin},
  {Strassmeier}, {Godoy-Rivera}, \& {Pinsonneault}}]{Metcalfe2019}
{Metcalfe}, T.~S., {Kochukhov}, O., {Ilyin}, I.~V., {Strassmeier}, K.~G.,
  {Godoy-Rivera}, D., \& {Pinsonneault}, M.~H. 2019, \apjl, 887, L38

\bibitem[{{Metcalfe} \& {van Saders}(2017)}]{Metcalfe2017}
{Metcalfe}, T.~S., \& {van Saders}, J. 2017, \solphys, 292, 126

\bibitem[{{Metcalfe} {et~al.}(2021){Metcalfe}, {van Saders}, {Basu}, {Buzasi},
  {Drake}, {Egeland}, {Huber}, {Saar}, {Stassun}, {Ball}, {Campante}, {Finley},
  {Kochukhov}, {Mathur}, {Reinhold}, {See}, {Baliunas}, \&
  {Soon}}]{Metcalfe2021}
{Metcalfe}, T.~S., {et~al.} 2021, \apj, 921, 122

\bibitem[{{Metcalfe} {et~al.}(2022){Metcalfe}, {Finley}, {Kochukhov}, {See},
  {Ayres}, {Stassun}, {van Saders}, {Clark}, {Godoy-Rivera}, {Ilyin},
  {Pinsonneault}, {Strassmeier}, \& {Petit}}]{Metcalfe2022}
---. 2022, \apjl, 933, L17

\bibitem[{{Petit} {et~al.}(2008){Petit}, {Dintrans}, {Solanki}, {Donati},
  {Auri{\`e}re}, {Ligni{\`e}res}, {Morin}, {Paletou}, {Ramirez Velez},
  {Catala}, \& {Fares}}]{Petit2008}
{Petit}, P., {et~al.} 2008, \mnras, 388, 80

\bibitem[{{R{\'e}ville} {et~al.}(2015){R{\'e}ville}, {Brun}, {Matt},
  {Strugarek}, \& {Pinto}}]{Reville2015}
{R{\'e}ville}, V., {Brun}, A.~S., {Matt}, S.~P., {Strugarek}, A., \& {Pinto},
  R.~F. 2015, \apj, 798, 116

\bibitem[{{Ryabchikova} {et~al.}(2015){Ryabchikova}, {Piskunov}, {Kurucz},
  {Stempels}, {Heiter}, {Pakhomov}, \& {Barklem}}]{Ryabchikova2015}
{Ryabchikova}, T., {Piskunov}, N., {Kurucz}, R.~L., {Stempels}, H.~C.,
  {Heiter}, U., {Pakhomov}, Y., \& {Barklem}, P.~S. 2015, \physscr, 90, 054005

\bibitem[{{Saar} \& {Osten}(1997)}]{Saar1997}
{Saar}, S.~H., \& {Osten}, R.~A. 1997, \mnras, 284, 803

\bibitem[{{Saar} \& {Testa}(2012)}]{SaarTesta2012}
{Saar}, S.~H., \& {Testa}, P. 2012, in Comparative Magnetic Minima:
  Characterizing Quiet Times in the Sun and Stars, ed. C.~H. {Mandrini} \&
  D.~F. {Webb}, Vol. 286, 335--345

\bibitem[{{Santos} {et~al.}(2021){Santos}, {Breton}, {Mathur}, \&
  {Garc{\'\i}a}}]{Santos2021}
{Santos}, A.~R.~G., {Breton}, S.~N., {Mathur}, S., \& {Garc{\'\i}a}, R.~A.
  2021, \apjs, 255, 17

\bibitem[{{Santos} {et~al.}(2019){Santos}, {Garc{\'\i}a}, {Mathur}, {Bugnet},
  {van Saders}, {Metcalfe}, {Simonian}, \& {Pinsonneault}}]{Santos2019}
{Santos}, A.~R.~G., {Garc{\'\i}a}, R.~A., {Mathur}, S., {Bugnet}, L., {van
  Saders}, J.~L., {Metcalfe}, T.~S., {Simonian}, G.~V.~A., \& {Pinsonneault},
  M.~H. 2019, \apjs, 244, 21

\bibitem[{{See} {et~al.}(2019){See}, {Matt}, {Folsom}, {Boro Saikia}, {Donati},
  {Fares}, {Finley}, {H{\'e}brard}, {Jardine}, {Jeffers}, {Lehmann}, {Marsden},
  {Mengel}, {Morin}, {Petit}, {Vidotto}, {Waite}, \& {BCool
  Collaboration}}]{See2019}
{See}, V., {et~al.} 2019, \apj, 876, 118

\bibitem[{{Stassun} {et~al.}(2017){Stassun}, {Collins}, \&
  {Gaudi}}]{Stassun2017}
{Stassun}, K.~G., {Collins}, K.~A., \& {Gaudi}, B.~S. 2017, \aj, 153, 136

\bibitem[{{Stassun} {et~al.}(2018){Stassun}, {Corsaro}, {Pepper}, \&
  {Gaudi}}]{Stassun2018}
{Stassun}, K.~G., {Corsaro}, E., {Pepper}, J.~A., \& {Gaudi}, B.~S. 2018, \aj,
  155, 22

\bibitem[{{Strassmeier} {et~al.}(2015){Strassmeier}, {Ilyin}, {J{\"a}rvinen},
  {Weber}, {Woche}, {Barnes}, {Bauer}, {Beckert}, {Bittner}, {Bredthauer},
  {Carroll}, {Denker}, {Dionies}, {DiVarano}, {D{\"o}scher}, {Fechner},
  {Feuerstein}, {Granzer}, {Hahn}, {Harnisch}, {Hofmann}, {Lesser}, {Paschke},
  {Pankratow}, {Plank}, {Pl{\"u}schke}, {Popow}, \&
  {Sablowski}}]{Strassmeier2015}
{Strassmeier}, K.~G., {et~al.} 2015, Astronomische Nachrichten, 336, 324

\bibitem[{{Tang} \& {Gai}(2011)}]{Tang2011}
{Tang}, Y.~K., \& {Gai}, N. 2011, \aap, 526, A35

\bibitem[{{Teixeira} {et~al.}(2009){Teixeira}, {Kjeldsen}, {Bedding}, {Bouchy},
  {Christensen-Dalsgaard}, {Cunha}, {Dall}, {Frandsen}, {Karoff}, {Monteiro},
  \& {Pijpers}}]{Teixeira2009}
{Teixeira}, T.~C., {et~al.} 2009, \aap, 494, 237

\bibitem[{{Torres} {et~al.}(2010){Torres}, {Andersen}, \&
  {Gim{\'e}nez}}]{Torres2010}
{Torres}, G., {Andersen}, J., \& {Gim{\'e}nez}, A. 2010, \aapr, 18, 67

\bibitem[{{Valenti} \& {Fischer}(2005)}]{ValentiFischer2005}
{Valenti}, J.~A., \& {Fischer}, D.~A. 2005, \apjs, 159, 141

\bibitem[{{van~Saders} {et~al.}(2016){van~Saders}, {Ceillier}, {Metcalfe},
  {Silva Aguirre}, {Pinsonneault}, {Garc{\'\i}a}, {Mathur}, \&
  {Davies}}]{vanSaders2016}
{van~Saders}, J.~L., {Ceillier}, T., {Metcalfe}, T.~S., {Silva Aguirre}, V.,
  {Pinsonneault}, M.~H., {Garc{\'\i}a}, R.~A., {Mathur}, S., \& {Davies}, G.~R.
  2016, \nat, 529, 181

\bibitem[{{van~Saders} \& {Pinsonneault}(2013)}]{vanSaders2013}
{van~Saders}, J.~L., \& {Pinsonneault}, M.~H. 2013, \apj, 776, 67

\bibitem[{{van~Saders} {et~al.}(2019){van~Saders}, {Pinsonneault}, \&
  {Barbieri}}]{vanSaders2019}
{van~Saders}, J.~L., {Pinsonneault}, M.~H., \& {Barbieri}, M. 2019, \apj, 872,
  128

\bibitem[{{von Braun} \& {Boyajian}(2017)}]{vonBraun2017}
{von Braun}, K., \& {Boyajian}, T. 2017, {Extrasolar Planets and Their Host
  Stars} (New York: Springer)

\bibitem[{{Wood}(2018)}]{Wood2018}
{Wood}, B.~E. 2018, in Journal of Physics Conference Series, Vol. 1100, Journal
  of Physics Conference Series, 012028

\bibitem[{{Wood} {et~al.}(2021){Wood}, {M{\"u}ller}, {Redfield}, {Konow},
  {Vannier}, {Linsky}, {Youngblood}, {Vidotto}, {Jardine},
  {Alvarado-G{\'o}mez}, \& {Drake}}]{Wood2021}
{Wood}, B.~E., {et~al.} 2021, \apj, 915, 37

\bibitem[{{Xiang} {et~al.}(2019){Xiang}, {Ting}, {Rix}, {Sandford}, {Buder},
  {Lind}, {Liu}, {Shi}, \& {Zhang}}]{Xiang2019}
{Xiang}, M., {et~al.} 2019, \apjs, 245, 34

\end{thebibliography}
\end{document}